\documentclass[a4paper,11pt]{article}

\usepackage{amssymb,latexsym}

\begin{document}
\topmargin -10mm \topmargin 0pt \oddsidemargin 0mm

\renewcommand{\thefootnote}{\fnsymbol{footnote}}
\newcommand{\nn}{\nonumber\\}
\begin{titlepage}

\vspace*{10mm}
\begin{center}
{\Large \bf Viscous Cosmology and Thermodynamics of Apparent
Horizon}

\vspace*{20mm}

{\large M. Akbar~\footnote{Email address: ak64bar@yahoo.com}
\footnote{Email address: makbar@camp.edu.pk}}\\
\vspace{8mm} { \em Centre for Advanced Mathematics and Physics\\
National University of Sciences and Technology\\
 Peshawar Road, Rawalpindi, Pakistan}

\end{center}

 \vspace{20mm}
 \centerline{{\bf{Abstract}}}
 \vspace{5mm}
It is shown that the differential form of Friedmann equations of FRW
universe can be recast as a similar form of the first law
,$T_{A}dS_{A} = dE + WdV$, of thermodynamics at the apparent horizon
of FRW universe filled with the viscous fluid. It is also shown that
the generalized second law of thermodynamics holds at the apparent
horizon of FRW universe and preserves dominant energy condition.

\end{titlepage}

\newpage
\renewcommand{\thefootnote}{\arabic{footnote}}
\setcounter{footnote}{0} \setcounter{page}{2}

\paragraph{Introduction:}
In the cosmological setting, one can associate Hawking temperature
and entropy with the apparent horizon analogous to the hawking
temperature and entropy associated with the black hole horizon
\cite{a9,a2,a15}. In the case of de Sitter space, the event horizon
matches with the apparent horizon of FRW universe with $k = 0$
however for more general cosmological models, the event horizon may
not exist but the apparent horizon associated with the Hawking
temperature and entropy always exists. The thermodynamic properties
associated with the apparent horizon of FRW universe filled with the
perfect fluid has been studied  by many authors (see for examples
\cite{a9, a2,jac, cai, hct, btz,a10}). The extension of this
connection between thermodynamics and gravity has also been carried
out in the braneworld cosmology \cite{b1,b2,b3,b4}. For a general
static spherically symmetric and stationary axisymmetric spacetimes,
It was shown that the Einstein field equations can be rewritten
\cite{ksp,Pad,PSP} as a first law of thermodynamics. More recently,
by considering a masslike function, it has been shown \cite{b5} that
the equilibrium thermodynamics should exist in the extended theories
of gravity and the Friedman equations in various theories of gravity
can be rewritten as a first law $TdS = dE$ at the apparent horizon
of FRW universe. More recently, Cai et al \cite{hct} has shown that
by employing Clausius relation, $\delta Q = TdS$, to the apparent
horizon of a FRW universe, they are able to derive the modified
Friedmann equation by using quantum corrected entropy-area
relation.\\
The cosmological models with perfect cosmic fluid has been studied
widely in literature however viscous cosmic fluid came much later in
the study of the universe \cite{min}. It has been revealed by Barrow
\cite{Barr}(also see \cite{Gio}) that the thermodynamic entropy
associated with bulk viscosity violates the dominant energy
condition and can decrease very well. This fact was resumed
\cite{Zim} in the framework of the solutions given in reference
\cite{Gio}. By taking into account the entropy of the sources only,
it has been deduced that the entropy of the bulk viscous sources
cannot decrease since it would violate the second law of
thermodynamics for large cosmic time values. However, in order to
state the generalized second law, one has to include both the
entropy connected with the sources and the entropy associated with
background geometry. The most of the study of the second law in the
cosmological context has been carried out at the event horizon (see
for examples \cite{Gio,Zim}) however the validity of second law
within the apparent horizon of FRW universe is of great important to
investigate. In this letter we will also investigate this issue. The
cosmological models with bulk viscosity have been studied in the
literature from various point of view (see for examples
\cite{Gio,bre,zimd}). \\The purpose of this letter is twofold. The
first is to show that the Friedmann equations of FRW universe
pertaining cosmic bulk viscosity can be rewritten as a similar form
of the first law of thermodynamics. The other is to discuss the
generalized second law within the apparent horizon of FRW
universe.\\
Let us start with a FRW universe of metric
\begin{equation}\label{1}
ds^{2}= -dt^{2}+ a^{2}(t)(\frac{dr^{2}}{1-kr^{2}}+r^{2}d\Omega^{2}),
\end{equation}
where $d\Omega^{2} = d\theta^{2} + sin^{2}\theta d\phi^{2}$ stands
for the line element of 2-dimensional unit sphere and the spatial
curvature constant $k = +1$, 0 and $-1$ represents a closed, flat
and open universe, respectively. The above metric (1) can be
rewritten in spherical form
\begin{equation}\label{2}
ds^{2}= h_{ab}dx^{a}dx^{b}+ \tilde{r}^{2}d\Omega^{2},
\end{equation}
where $\tilde{r} = a(t)r$, $x^{0} = t$, $x^{1}= r$ and $h_{ab} =
diag(-1,  \frac{a^{2}}{1-kr^{2}})$. One can work out the dynamical
apparent horizon from the relation
$h^{ab}\partial_{a}\tilde{r}\partial_{b}\tilde{r} = 0$ which turns
out $\frac{1}{\tilde{r}^{2}_{A}} = H^{2}+k/a^{2}$, where $H \equiv
\frac{\dot{a}}{a}$ is the Hubble parameter and the dots denote
derivatives with respect to the cosmic time. let the universe be
filled with a viscous fluid and $u^{\mu}=(u^{0}, u^{i})$ is the
four-velocity of the fluid. In comoving coordinates $u^{0}=1$ and
$u^{i}=0$. Define $h_{\mu\nu} = g_{\mu\nu}+ u_{\mu}u_{\nu}$,
$w_{\mu\nu}=h^{\alpha}_{\mu}h^{\beta}_{\nu}u_{[\alpha;\beta]}$ and
$\theta_{\mu\nu}=h^{\alpha}_{\mu}h^{\beta}_{\nu}u_{(\alpha;\beta)}$
as a projection tensor, rotation tensor and expansion tensor,
respectively. The scalar expansion is $\theta \equiv \theta
^{\mu}_{\mu}= u^{\mu}_{;\mu}$ and
$\sigma_{\mu\nu}=\theta_{\mu\nu}-\frac{1}{3}h_{\mu\nu}\theta$ is the
shear tensor. Taking into consideration of metric (1),
$u^{\mu}_{;\nu} = h^{\mu}_{\nu}\frac{\dot{a}}{a}$ which implies that
the rotation and shear tensors vanish, i.e. $w_{\mu\nu} =
\sigma_{\mu\nu} = 0$. While the scalar expansion is $\theta =
3\frac{\dot{a}}{a} = 3H$. Hence the energy-momentum tensor of
viscous fluid in the background of metric (1) finally becomes
\cite{bre}
\begin{equation}\label{5}
T_{\mu\nu}=(\rho +P-\zeta \theta)u_{\mu}u_{\nu}+(P-\zeta
\theta)g_{\mu\nu},
\end{equation}
where $\rho$, $P$ and $\zeta$ are the energy density, thermodynamic
pressure and bulk viscosity respectively. The components of
energy-momentum tensor are $T_{00} = \rho $, $T_{0i}=0$ and $T_{ij}
= (P-\zeta \theta)g_{ij}$. Therefore, the total effect of the bulk
viscosity is to reduce the pressure $P$ of the perfect fluid by an
amount $\zeta \theta$, so that the effective pressure of the viscous
fluid turns out to be $\tilde{P} = P -\zeta \theta$. The energy
conservation $T^{\mu\nu}_{;\nu} = 0$, yields $\dot{\rho} + 3H(\rho +
\tilde{P}) = 0$. Now we turn to discuss the thermodynamic
interpretation of Friedman equations at the apparent horizon of FWR
universe. For this purpose, we first define the surface gravity
$\kappa =
\frac{1}{2\sqrt{-h}}\partial_{a}(\sqrt{-h}h^{ab}\partial_{b}\tilde{r})$
at the apparent of FRW universe. By utilizing above relation, one
can easily find that the surface gravity at the apparent horizon of
FRW universe yields
\begin{equation}
\kappa =
\frac{-1}{\tilde{r}_{A}}(1-\frac{\dot{\tilde{r}}_{A}}{2H\tilde{r}_{A}}).
\end{equation}
When $\frac{\dot{\tilde{r}}_{A}}{2H\tilde{r}_{A}}\leq 1$ implies the
surface gravity $\kappa \leq 0$ which leads to the temperature $T =
\kappa / 2\pi \leq 0 $. However, in reference \cite{a9}, an
approximation, $\frac{\dot{\tilde{r}}_{A}}{2H\tilde{r}_{A}}\ll 1$,
has been used while determining horizon temperature. They found $T =
\frac{1}{2\pi \tilde{r}_{A}}$ by approximating surface gravity
$|\kappa| = \frac{1}{\tilde{r}_{A}}$. Now we turn to define energy
of the universe within the apparent horizon. We take total matter
energy $E = V\rho$ inside a sphere of radius $\tilde{r}$ which is
also the Miser-Sharp energy \cite{sharp}, $E = \frac{\tilde{r}}{2
}(1-h^{ab}\partial_{a}\tilde{r}\partial_{b}\tilde{r})$, within the
apparent horizon. Since at the apparent horizon
$(h^{ab}\partial_{a}\tilde{r}\partial_{b}\tilde{r} = 0)$, so the
Mizner-Sharp energy is in fact total matter energy inside the sphere
of radius $\tilde{r}_{A}$ and is given by
\begin{equation}
E = V\rho.
\end{equation}
The entropy $S = A /4$ is the horizon entropy, where $A = 4\pi
\tilde{r}_{A}^{2}$ is the horizon area. Note that we use the units
$\hbar = c = G = \kappa_{B} = 1$. From the Einstein field equations
$G_{\mu\nu} = 8\pi  T_{\mu\nu}$, the Friedman equations for the
viscous fluid of stress-energy tensor (3) can be written as
\begin{equation}
H^{2}+ \frac{k}{a^{2}} = \frac{8\pi }{3}\rho,
\end{equation}
\begin{equation}
\dot{H}-\frac{k}{a^{2}} = -4\pi  (\rho + \tilde{P}).
\end{equation}
In terms of the apparent horizon, the Friedman equation (6) can be
rewritten as
\begin{equation}
\frac{1}{\tilde{r}^{2}_{A}} = \frac{8\pi }{3}\rho.
\end{equation}
Let the apparent horizon surface acts as the boundary of the thermal
system of the FRW universe. In general, the apparent horizon is not
constant but changes with time. Let $d\tilde{r}_{A}$ be an
infinitesimal change in radius of the apparent horizon in time
interval dt. This change in the apparent horizon will cause a small
change $dV$ in the volume $V$ of the universe. This constitutes two
spherical system of space-time with radii $\tilde{r}_{A}$ and
$\tilde{r}_{A} + d\tilde{r}_{A}$ with a common source of fluid with
non-zero pressure and energy density near horizon. Each space-time
constituting a thermal system and satisfying Einstein's equations,
differs infinitesimally in the extensive variables volume, energy
and entropy by $dV$, $dE$ and $dS$, respectively, while having the
same values of intensive variables temperature $T$ and pressure
$\bar{P}$. Thus for these two thermal states of the space-time,
there must exist a certain relation connecting these thermodynamic
quantities. To establish connection among the thermal quantities and
Friedman equations, we differentiate equation (8) which implies
\begin{equation}
\frac{1}{\tilde{r}_{A}^{3}}d\tilde{r}_{A} = 4 \pi (\rho +
\tilde{P})Hdt
\end{equation}
Multiplying both sides of this equation with a factor $
\tilde{r}_{A}^{3}(1-\frac{\dot{\tilde{r}}_{A}}{2H\tilde{r}_{A}})$,
one can rewrite this equation in a form
\begin{equation}
\frac{\kappa}{2\pi}\frac {d(4\pi \tilde{r}_{A}^{2})}{4} = - 4\pi
\tilde{r}_{A}^{2}(\rho + P)H(1 -
\frac{\dot{\tilde{r}}_{A}}{2H\tilde{r}_{A}})dt
\end{equation}
One can recognize that the quantities $\frac{\kappa}{2\pi}$ and
$\frac{4\pi \tilde{r}_{A}^{2}}{4}$ are the temperature $T$ and
entropy $S$ respectively. Therefore the above equation can be
rewritten as
\begin{equation}
T dS = - 4\pi \tilde{r}_{A}^{3}(\rho + \tilde{P})H(1 -
\frac{\dot{\tilde{r}}_{A}}{2H\tilde{r}_{A}})dt
\end{equation}
Now we turn to the total matter energy (5) and taking differential
of it, one gets
\begin{equation}
dE = 4\pi \tilde{r}_{A}^{2}\rho d\tilde{r}_{A} - 4\pi
\tilde{r}_{A}^{3}(\rho + \tilde{P})Hdt.
\end{equation}
Using equations (11) and (12) one yields
\begin{equation}
dE = TdS + WdV,
\end{equation}
where $W = \frac{1}{2}(\rho - \tilde{P})$ is the work density. The
above equation is the unified first law \cite{Hay} of relativistic
thermodynamics. Note that the above thermal identity is obtained
from the Friedman equation with viscous fluid together with the
characteristics of the apparent horizon, while the author of
reference \cite{Hay} studied the thermodynamics of trapping horizon
of dynamical black hole. It has been found that the entropy
associated with apparent horizon is proportional to the horizon area
which is originally initiated from the black hole horizon entropy
that satisfies the so-called area formula \cite{wald}. In the case
of perfect fluid, $\zeta = 0$, one can compare the above thermal
identity with the standard form of the first law, $dE = TdS -PdV$,
of thermodynamics. In fact, the negative pressure term $-P$ in the
first law is taken the place of the work density $W$. Notice that
for pure de Sitter spacetime, $\rho = -P$, then one acquires the
standard form $dE = TdS-PdV$. Note that we are considering the
universe as a thermal object with apparent horizon as its boundary.
So it useful to define other thermodynamic quantities like Enthalpy
$H$ and Gibbs free energy $G$. The Enthalpy is defined by the
equation $H = U + PV$, where $U = E$ is the total internal energy
which is taken to be the total matter energy of the universe bounded
by the apparent horizon. So the Enthalpy of the universe bounded by
the apparent horizon can be written in terms of apparent radius $H =
\frac{4\pi}{3}(\rho + P)\tilde{r}_{A}^{3}$. So the heat capacity of
the universe enveloped by the apparent horizon at constant pressure
of the perfect is defined as
\begin{equation}
C_{P} = (\frac{\partial H}{\partial T})_{P} = (\frac{\partial
H}{\partial \tilde{r}_{A}}\frac{\partial \tilde{r}_{A}}{\partial
T})_{P},
\end{equation}
which yields $C_{P} = 4\pi \tilde{r}_{A}^{2}(\rho + P)
(\frac{\partial \tilde{r}_{A}}{\partial T})_{P}$. In general the
temperature, $T = \kappa / 2\pi \equiv \frac{-1}{2\pi
\tilde{r}_{A}}(1-\frac{\dot{\tilde{r}}_{A}}{2H\tilde{r}_{A}})$.
However, if one considers the approximation,
$\frac{\dot{\tilde{r}}_{A}}{2H\tilde{r}_{A}} \ll 1$ (also see
\cite{a9, b1}), the temperature $T = \mid\kappa\mid / 2\pi \equiv
1/2\pi \tilde{r}_{A}$ which implies that the heat capacity of the
universe is negative provided the dominant energy condition holds.
On the other hand, if one consider general temperature $T =
\frac{-1}{2\pi
\tilde{r}_{A}}(1-\frac{\dot{\tilde{r}}_{A}}{2H\tilde{r}_{A}})$
associated with the apparent horizon which implies the heat capacity
of the universe is positive definite provided
$\frac{\dot{\tilde{r}}_{A}}{H\tilde{r}_{A}}\leq 1$. It is also
interesting to note that the total matter energy $E = V \rho$,
entropy $S = A / 4 $ and temperature $T = 1/2\pi \tilde{r}_{A}$
satisfy the relation $E = TS$ .
\paragraph{Generalized Second Law:} Recently a lot of attention has
been granted to the generalized second law of thermodynamics in the
accelerating universe handled by the dark energy \cite{jia}. Using a
particular model for dark energy, the generalized second law has
been studied in reference \cite{bin} with the boundaries of universe
enveloped by the apparent as well event horizon. It is important to
investigate the generalized second law as defined in the region of
the universe bounded by the apparent horizon in more general
context.\\ Let us now consider a region of FRW universe bounded by
the apparent horizon filled by a perfect fluid of energy density
$\rho$ and pressure $P$. We assume that the region bounded by the
apparent horizon acts as a thermal system with boundary defined by
the apparent horizon. Since the apparent horizon is not constant but
varies with time. As the apparent radius changes , the volume
enveloped by the apparent horizon will also change, however the
thermal system bounded by the apparent horizon remains in
equilibrium when it moves from one state to another so that the
temperature of the system must be uniform and the same as the
temperature of its surroundings. This requires that the temperature
of total energy inside the apparent horizon should be in equilibrium
with the temperature associated with the apparent horizon because we
are not considering the flow of energy through the horizon. \\
In order to state that a given cosmological solution
satisfies/violates the second law of thermodynamics one should
define first what is the generalization of the second law to the
case of FRW universe. It should include both the entropy due to
sources and the entropy associated with the background geometry. So
the generalized second law can be expressed as
\begin{equation}
\dot{S}_{A} + \dot{S}_{m} \geq  0,
\end{equation}
where $S_{A} = A/4G$ is the Bekenstein-Hawking entropy associated
with the apparent horizon of FRW universe and $S_{m}$ is the entropy
due to the matter sources inside the apparent horizon. From the
Friedmann equations (6) and (7), it can easily be shown that
$\dot{\tilde{r}}_{A} = 4\pi  \tilde{r}_{A}^{3}H(\rho + P)\geq 0$
provided the universe driven by the source of perfect fluid
preserves dominant energy condition. Since we are assuming the
region of a FRW universe enveloped by the apparent horizon as
thermal system so that the change of energy $dE$ of the universe
from one state to another must be connected through the Friedmann
equations together with the characteristics of the apparent horizon.
It has been shown \cite{b1,a15,b2} that the Friedmann equations at
apparent horizon of FRW universe filled with perfect fluid satisfied
the thermal identity $dE = T_{A}dS_{A} + WdV$ instead of satisfying
the standard first law $T_{A}dS_{A} = dE + PdV$, where $T_{A}$ is
the temperature associated with the apparent horizon. Since we are
assuming the local equilibrium so that the change of matter energy
$dE_{m}$ satisfies the first law of thermodynamics, $dE_{m} =
T_{A}dS_{m} - PdV$, where $T_{A} = |\kappa|/2\pi$ is the temperature
associated with the apparent horizon and is given by
\begin{equation}
T = \frac{1}{2\pi
\tilde{r}_{A}}(1-\frac{\dot{\tilde{r}}_{A}}{2H\tilde{r}_{A}}),
\end{equation}
where $\frac{\dot{\tilde{r}}_{A}}{2H\tilde{r}_{A}}\leq 1$ ensures
that the temperature is positive. From the first law, $dE_{m} =
T_{A}dS_{m} - PdV$, one can obtain
\begin{equation}
T_{A}\dot{S}_{m} = 4\pi \tilde{r}_{A}^{2}\dot{\tilde{r}}_{A}(\rho +
P)- 4\pi \tilde{r}_{A}^{3}H(\rho + P).
\end{equation}
Now we turn to find out the entropy change $\dot{S}_{H}$ associated
with the apparent horizon of FRW universe which yields
\begin{equation}
T_{A}\dot{S}_{A} = 4\pi \tilde{r}_{A}^{3}H(\rho + P)-2\pi
\tilde{r}_{A}^{2}(\rho + P)\dot{\tilde{r}}_{A}.
\end{equation}
Adding equations (17) and (18), one gets the expression for the
generalized second law
\begin{equation}
T_{A}(\dot{S}_{m}+\dot{S}_{H}) = 2 \pi
\tilde{r}_{A}^{2}\dot{\tilde{r}}_{A}(\rho + P).
\end{equation}
It is obvious from the above equation that
$\dot{S}_{m}+\dot{S}_{H}\geq 0$ implies that the generalized second
law holds provided $\rho + P \geq 0$.
\paragraph{Conclusion:}
It is shown that the differential form of Friedmann equations of FRW
universe filled with a viscous fluid can be rewritten as a similar
form of the first law ,$TdS = dE + WdV$, of thermodynamics at the
apparent horizon of FRW universe. It can easily be seen that the
similar identity also holds for the perfect fluid when bulk
viscosity of the fluid is zero. It is also shown that the heat
capacity of the universe filled with perfect fluid is negative if
one considers the approximate temperature $T = |\kappa| / 2\pi=
1/2\pi \tilde{r}_{A}$. However if one utilizes  the general
expression $T = \frac{-1}{2\pi
\tilde{r}_{A}}(1-\frac{\dot{\tilde{r}}_{A}}{2H\tilde{r}_{A}})$, the
heat capacity of the universe is positive definite provided
$\frac{\dot{\tilde{r}}_{A}}{H\tilde{r}_{A}} \leq 1$ which implies
that the universe within the apparent horizon of FRW universe is
thermodynamically stable. We also verify that the generalized second
law of thermodynamics at the apparent horizon of FRW universe holds
provided that the matter source satisfies the dominant energy
condition. It is interesting to investigate the generalized second
law for the extended theories of gravity in a region bounded by the
apparent horizon of FRW universe. The work in this respect is under
progress.

\section*{Acknowledgments} I would like to thank Rong-Gen Cai for
his useful comments. The work is supported by a grant from National
university of Sciences and Technology.


\end{document}